\documentclass[12pt]{article}
\pdfoutput=1
\usepackage[pdftex]{graphicx}
\usepackage{amssymb}
\usepackage{amsmath}
\usepackage{bm}
\usepackage{enumerate}

\begin{document}

\begin{titlepage}
	\vskip 2cm
	\begin{center}
		\Large{{\bf On a new relation between entanglement\\and geometry from M(atrix) theory}}
	\end{center}

	\vskip 2cm
	\begin{center}
		{Vatche Sahakian\footnote{\tt{sahakian@hmc.edu}}}\\
	\end{center}
	\vskip 12pt
	\centerline{\sl Harvey Mudd College}
	\centerline{\sl Physics Department, 241 Platt Blvd.}
	\centerline{\sl Claremont CA 91711 USA}

	\vskip 1cm
	\begin{abstract}
		In the context of Matrix/light-cone gauge M-theory, we develop a new approach for computing quantum entanglement between a probe gravitating in the vicinity of a source mass and the source mass. We demonstrate that this entanglement is related to the gravitational potential energy between the two objects. We then show that the Von Neumann entropy is a function of two derivatives of the gravitational potential. We conjecture a relation between the entropy and the local Riemann tensor sampled by the probe, establishing a general scheme to relate entropy to local geometric data. This relation connects the rate of change, rotation, and twist of a small volume element at the probe's location to the quantum entanglement of the probe with the source. 
	\end{abstract}
\end{titlepage}

\newpage \setcounter{page}{1}

\section{Introduction and highlights}
\label{sub:intro}

Various relations between quantum information and spacetime geometry seem to hint at the need for a fundamental rethinking of gravity. In this program, the general theme appears to be that gravity is an emergent phenomenon; and that underlying microscopic quantum degrees of freedom weave -- through quantum entanglement -- a fabric that we effectively perceive as space. In this note, we want to analyze these ideas in the context of Matrix theory, a non-perturbative formulation of string theory and quantum gravity~\cite{Banks:1996vh}. We will consider a simple setup where a massive source pulls gravitationally on a probe; and where it is well-known that the effective quantum potential that arises from Matrix theory matches exactly with the expected gravitational potential that the probe experiences in light-cone gauge M-theory~\cite{Kabat:1997im}. This effective potential arises from integrating out fast off-diagonal matrix modes that correspond to strings stretched between the two objects. In this work, we add the slower diagonal excitations and derive their quantum effective potential. We then demonstrate that the quantum vacuum of these modes is an entangled state in such a way that the entanglement entropy between source and probe is generally a function of derivatives of their gravitational potential. We compute the Von Neumann entropy and, based on the result we obtain, we conjecture a relation between the entropy and the local Riemann tensor sampled by the probe.
Essentially, this entanglement entropy is shown to be directly related to local tidal forces. This connects the entropy to the rate of change, rotation, and twist of a small volume element at the location of the probe. The setup is reminiscent of entropy-area relations, except the statement we obtain is local. 

In the first section, we describe the setup and outline the computation of the entanglement entropy. In the second section, we present a conjecture relating this entropy to local geometry. The Conclusion section discusses the more general implications of these results and future directions.

\section{Quantum entanglement and gravity}
\label{sub:arelation}

Matrix theory is $0+1$ dimensional $U(N)$ Super Yang-Mills (SYM) theory that is purported to be dual to
light-cone gauge M-theory. The rank of the gauge group $N$ maps onto light-cone momentum in M-theory. 
Our starting point is the Matrix theory action in the background field gauge\footnote{We will try to follow, as much as possible, the notation and conventions used in~\cite{Kabat:1997im} and~\cite{Sugiyama:2002bw}.}
\begin{eqnarray}
	S&=&\frac{1}{g^2_{\mathrm{YM}}} \int dt\, \mbox{Tr} \left[
	D_t X^i\, D_t X^i +\frac{1}{2} [X^i,X^j]^2 -(\partial_t A_0-i[X^i_{\mathrm{bg}},X^i])^2 \right. \nonumber \\
	&+& \left. i\, \Psi_\alpha D_t \Psi_\alpha - \Psi_\alpha \Gamma^i_{\alpha\beta} [X^i,\Psi_\beta] + i \overline{G} \partial_t D_t G+\overline{G}[X^i_{\mathrm{bg}},[X^i,G]]
	\right]\ .\label{eq:matrixtheory}
\end{eqnarray}
All fields are in the adjoint of $U(N)$, and the spinor fields $\Psi_\alpha$ are $10$ dimensional Majorana-Weyl.
The last term in the first line is a gauge fixing term for the condition
\begin{equation}
	\partial_t A_0-i[X^i_{\mathrm{bg}},X^i] = 0\ ,
\end{equation}
and $G$ is a matrix of Faddeev-Popov ghosts. The Yang-Mills coupling is given by $g^2_{\mathrm{YM}}=2\,R$ where $R$ is the radius of the M-theory light-cone circle. 
We work in string units, $\ell_s=1$.
We take the background as
\begin{equation}
	X^i_{\mathrm{bg}} = \left(
	\begin{array}{cc}
		\overline{X}^i_1(t) & 0 \\
		0 & \overline{X}^i_2(t)
	\end{array}
	\right)
\end{equation}
with all other fields vanishing.
This is a block diagonal configuration with $\overline{X}^i_1$ being an $N_1\times N_1$ matrix, and $\overline{X}^i_2$ being an $N_2\times N_2$ matrix; we have $N=N_1+N_2$. In M-theory language, $\overline{X}^i_1$ is to represent an object that carries $N_1$ units of light-cone momentum -- such as a spherical mass or a graviton; while $\overline{X}^i_2$ represents another object with $N_2$ units of light-cone momentum. We then want to write down an effective action by perturbing this background by
\begin{equation}
	\begin{array}{lll}
		A_0 = \left(
		\begin{array}{cc}
			a_1(t) & a(t) \\
			\overline{a}(t) & a_2(t)
		\end{array}
		\right) &
		X^i = X^i_{\mathrm{bg}} + \left(
		\begin{array}{cc}
			x_1^i(t) & x^i(t) \\
			{x}^{i\,\dagger}(t) & x_2^i(t)
		\end{array}
		\right) & 
		\Psi_\alpha = \left(
		\begin{array}{cc}
			\psi_{1\alpha}(t) & \psi_\alpha(t) \\
			{\psi^\dagger}_\alpha(t) & \psi_{2\alpha}(t)
		\end{array}
		\right)
	\end{array}\label{eq:perturbations}\ .
\end{equation}
The centers of mass of the two background objects are given by
\begin{equation}
	\overline{x}_{1,2}^i \equiv \frac{\mbox{Tr}\, \overline{X}_{1,2}^i}{N_{1,2}}
\end{equation}
while the size of each object might naturally be represented by the second moments
\begin{equation}
	{R}_{1,2}^2 \equiv \frac{\mbox{Tr}\, (\overline{X}_{1,2}^i)^2}{N_{1,2}} - (\overline{x}_{1,2}^i)^2\ .
\end{equation}
We assume that the two background objects are widely separated from each other so that their gravitational potential energy is small compared to their kinetic energies. We also assume that their sizes are much smaller than the distance between them. In this regime, the off-diagonal perturbations in~(\ref{eq:perturbations}) are heavy or high frequency modes. One can then integrate them out and discovers that, for large $N_{1,2}$ and while setting all diagonal perturbations to zero, the resulting effective potential for the background variables $\overline{X}_1^i$ and $\overline{X}_2^i$ agrees with the Newtonian gravitational potential between the two objects in light-cone gauge M-theory~\cite{Kabat:1997im}. This is a remarkable result in support of the Matrix theory-M theory correspondence. 

Our task is to add to this computation the lighter, slower perturbations on the diagonal: the $x_{1,2}$'s, $a_{1,2}$'s, and $\psi_{1,2}$'s. We then want to write the effective potential for the $x_{1,2}$ and $\psi_{1,2}$ after the fast modes are integrated out. We write the effective potential, after integrating out the heavy off-diagonal modes, as
\begin{equation}\label{eq:S0SV}
	S_{eff} = S_0+S_V
\end{equation}
where the first term $S_0$ comes from the part of the action that does {\em not} involve the off-diagonal perturbations and takes the form
\begin{eqnarray}
	S_0 &=& \int dt\, \mbox{Tr}\left(
	(\partial_t x_1^i)^2 + (\partial_t x_2^i)^2 +[x_1^i,x_1^j][\overline{X}_1^i,\overline{X}_1^j]
	+[x_1^i,\overline{X}_1^j][x_1^i,\overline{X}_1^j]-[x_1^i,\overline{X}_1^j][x_1^j,\overline{X}_1^i] \right. \nonumber \\
	&+& \left. [x_2^i,x_2^j][\overline{X}_2^i,\overline{X}_2^j] +[x_2^i,\overline{X}_2^j][x_2^i,\overline{X}_2^j]-[x_2^i,\overline{X}_2^j][x_2^j,\overline{X}_2^i] \right. \nonumber \\
	&-& \left.(\partial_t a_1)^2-(\partial_t a_2)^2+2\,i\, (\partial_t a_1)\, [\overline{X}^i_1,x^i_1] + 2\,i\, (\partial_t a_2)\, [\overline{X}^i_2,x^i_2] \right. \nonumber \\
	&+& \left. i\, \psi_{1\alpha} \partial_t \psi_{1\alpha}+ i\, \psi_{2\alpha} \partial_t \psi_{2\alpha} + \psi_{1\alpha} \Gamma^i_{\alpha\beta} [\psi_{1\beta},x_1^i]+ \psi_{2\alpha} \Gamma^i_{\alpha\beta} [\psi_{2\beta},x_2^i]
	\right)\ .\label{eq:S0}
\end{eqnarray}
An important observation here is that there are no $x_1$-$x_2$ couplings in $S_0$; hence, the coupling between the two objects, and thus any entanglement between them, can come only from $S_V$. Furthermore, there are no $\psi_1$-$\psi_2$ coupling terms in $S_0$; nor will there be any in $S_V$: to leading order in small perturbations, given the action's quadratic form in the fermions, there is no entanglement to be considered between the fermionic diagonal modes.

The second piece of~(\ref{eq:S0SV}), $S_V$, involves the off-diagonal perturbations that can be integrated out in the regime of interest. The computation of $S_V$ proceeds as in~\cite{Kabat:1997im} where the diagonal perturbations were set to zero, except now $\overline{X}^i_1$ and $\overline{X}^i_2$ are now shifted by $x_1^i$ and $x_2^i$; we get from the ground state energy of the oscillators~\cite{Kabat:1997im}
\begin{equation}\label{eq:Sv}
	S_V = -\int dt\, \left(\mbox{Tr'} \sqrt{M_{0b}+M_{1b}}-\frac{1}{2} \mbox{Tr'} \sqrt{M_{0f}+M_{1f}}
	-2\,\mbox{Tr'} \sqrt{M_g}\right)\ ,
\end{equation}
where we define the `mass matrices' along~\cite{Kabat:1997im}: from the bosonic sector involving $x$ and $a$, we have
\begin{equation}
	M_{0b} = \sum_i K^{i\,2}\otimes \bm{1}_{10\times 10}\ \ \ ,\ \ \ 
	M_{1b} = \left(
	\begin{array}{cc}
		\bm{0} & -2\, \partial_t K^j \\
		2\,\partial_t K^i & 2\,[K^i,K^j]
	\end{array}
	\right)\ ;
\end{equation}
from the fermionic sector involving $\psi$, we have
\begin{equation}
	M_{0f} = \sum_i K^{i\,2}\otimes \bm{1}_{16\times 16}\ \ \ ,\ \ \ 
	M_{1f} = i\,\partial_t K^i \otimes \Gamma^i+\frac{1}{2}[K^i,K^j]\otimes \Gamma^{ij}\ ;
\end{equation}
and from the ghost sector, we get
\begin{equation}
	M_g = \sum_i K^{i\,2}\ .
\end{equation}
In these expressions, we have defined the matrix
\begin{equation}\label{eq:K}
	K^i = (\overline{X}^i_1+x_1^i) \otimes \bm{1}_{N_2\times N_2} - \bm{1}_{N_1\times N_1}\otimes (\overline{X}^{i\,T}_2+x_2^{i\,T})
\end{equation}
In~(\ref{eq:Sv}), $\mbox{Tr'}$ corresponds to tracing over both group and Lorentz spaces.
Throughout, we are assuming, as in~\cite{Kabat:1997im}, that the background satisfies the equations of motion, and hence all terms linear in the perturbations should be dropped. Hence, it is implicit in~(\ref{eq:Sv}) that we drop linear terms in $x_1^i$ and $x_2^i$ once the expression is expanded further. Given the similarities between~(\ref{eq:Sv}) and the result in~\cite{Kabat:1997im}, with the only modification coming from the shifts by $x_1^i$ and $x_2^i$ in~(\ref{eq:K}), the computations proceed along similar steps: we write the square root of the matrices using a Dyson perturbation series in $M_{1b}$ and $M_{1f}$, where $M_{1b}$ and $M_{1f}$ are smaller than $M_{0b}$ and $M_{0f}$. The zeroth order corresponds to zero point energy and cancels by supersymmetry once we include the contribution from the ghosts ($M_g$ only contributes to zeroth order); the cancellations carry over to linear, quadratic, and third order in $M_1$. The first non-zero contribution arises at fourth order and we get
\begin{eqnarray}
	&& S_V = \frac{1}{2\sqrt{\pi}}\mbox{Tr} \int dt\, \left(
		\int_0^\infty\int_0^\infty\int_0^\infty\int_0^\infty\int_0^\infty \frac{d\tau_1\,d\tau_2\,d\tau_3\,d\tau_4\,d\tau_5}{(\tau_1+\tau_2+\tau_3+\tau_4+\tau_5)^{3/2}} 
	\right. \nonumber \\
	&& \left. e^{-(\tau_1+\tau_2+\tau_3+\tau_4+\tau_5) M_{0b}} \mbox{Tr}_L \left[M_{1b}(\tau_2+\tau_3+\tau_4+\tau_5)M_{1b}(\tau_3+\tau_4+\tau_5)M_{1b}(\tau_4+\tau_5)M_{1b}(\tau_5)\right] \right. \nonumber \\
	&&- \left. \frac{1}{2}e^{-(\tau_1+\tau_2+\tau_3+\tau_4+\tau_5) M_{0f}} \mbox{Tr}_L \left[M_{1f}(\tau_2+\tau_3+\tau_4+\tau_5)M_{1f}(\tau_3+\tau_4+\tau_5)M_{1f}(\tau_4+\tau_5)M_{1f}(\tau_5)\right]
	\right) \nonumber \\ \label{eq:dyson}
\end{eqnarray}
where we defined
\begin{equation}
	M_1(\tau) \equiv e^{\tau\, M_0}\, M_1\, e^{-\tau\, M_0}\ .
\end{equation}
$\mbox{Tr}_L$ involves tracing over Lorentz space, while $\mbox{Tr}$ refers to tracing over group space as usual. Let us then write 
\begin{equation}
	K^i = \overline{K}^i+\Delta K^i
\end{equation}
where we define 
\begin{equation}
	\overline{K}^i \equiv \overline{X}^i_1 \otimes \bm{1}_{N_2\times N_2} - \bm{1}_{N_1\times N_1}\otimes \overline{X}^{i\,T}
\end{equation}
so that all diagonal perturbations are in the $\Delta K^i$ matrix. To proceed further, we will focus onto a subsector of diagonal perturbations that perturb the location of the centers of masses of the two objects. We write
\begin{equation}\label{eq:specialperturbations}
	x_1^i = \varepsilon_1^i \bm{1}_{N_1\times N_1}\ \ \ ,\ \ \ 
	x_2^i = \varepsilon_2^i \bm{1}_{N_2\times N_2}\ ,
\end{equation}
where $\varepsilon_1^i$ and $\varepsilon_2^i$ are now the small perturbations associated with blocks 1 and 2 respectively.
Beyond being a physically natural choice, these perturbations also decouple from other perturbations as they drop out of the commutators appearing in~(\ref{eq:matrixtheory}). This means that truncating to this sector of perturbations is mathematically consistent. The first part of the action given by $S_0$ in~(\ref{eq:S0}) then becomes
\begin{equation}\label{eq:S02}
	S_0 = \int dt\,\left( N_1 (\partial_t \varepsilon^i_1)^2 + N_2 (\partial_t \varepsilon^i_2)^2 + \mbox{Tr} \left[ - (\partial_t a_1)^2 - (\partial_t a_2)^2 + i\, \psi_{1\alpha} \partial_t \psi_{1\alpha}+ i\, \psi_{2\alpha} \partial_t \psi_{2\alpha}\right]\right)
\end{equation}
We also have
\begin{equation}
	\Delta K^i = (\varepsilon^i_1 - \varepsilon^i_2)\  \bm{1}_{N_1\times N_1} \otimes \bm{1}_{N_2\times N_2}\ .
\end{equation}
We then get
\begin{equation}
	K^{i\,2} = (\overline{K}^i+\Delta K^i)^2 = \overline{K}^{i\,2} + 2\, (\varepsilon^i_1 - \varepsilon^i_2)\,\overline{K}^i +  (\varepsilon^i_1 - \varepsilon^i_2)^2 \bm{1}_{N_1\times N_1} \otimes \bm{1}_{N_2\times N_2}\ .
\end{equation}
Assuming that the size of each object $R_1$ and $R_2$ is much smaller than the separation distance between them, the eigenvalues of $\overline{K}^{i\,2}$ scale as $r^2$ where we define
the relative position vector between the centers of mass of the two objects as
\begin{equation}
	r^i = \overline{x}_1^i-\overline{x}_2^i\ .
\end{equation}
More generally, we expect that
\begin{equation}
	\overline{K}^i = r^i\ \bm{1}_{N_1\times N_1} \otimes \bm{1}_{N_2\times N_2} + \kappa^i\label{eq:kappa}
\end{equation}
where $\kappa^i$ is a matrix whose entries scale at most as $R_1^2$ and $R_2^2$, the characteristic sizes of the two objects -- independent of the distance $r$ separating them. As long as $R_{1,2} \ll r$, 
we can then approximately write
\begin{equation}
	K^{i\,2} \simeq r^2 + 2 (\varepsilon^i_1 - \varepsilon^i_2)\,r^i + r^2 = (r^i + \varepsilon^i_1 - \varepsilon^i_2)^2
\end{equation}
which is large, scaling as $r^2$ with large $r$.
Looking back at~(\ref{eq:dyson}), focus first on the exponential factor in the integrand. Whether for bosons or fermions, we have a structure of the form
\begin{equation}
	e^{-(\tau_1+\tau_2+\tau_3+\tau_4+\tau_5) K^{i\,2}}\ .
\end{equation}
For large $r$, this implies that the predominant contribution to the integral in~(\ref{eq:dyson}) comes from the region where the $\tau$'s are zero. As a result, we can approximately write, as in~\cite{Kabat:1997im},
\begin{equation}
	M_1(\tau) = e^{\tau\, M_0}\, M_1\, e^{-\tau\, M_0} \simeq M_1\ .
\end{equation}
This leads to a very similar expression to the effective Newtonian potential computed in~\cite{Kabat:1997im}, now given by
\begin{eqnarray}
	S_V &=& \int dt\, \frac{5}{128\, ((\overline{x}_1^i-\overline{x}_2^i + \varepsilon^i_1 - \varepsilon^i_2)^2)^{7/2}} \mbox{Tr} \left[
		8\, F^\mu_{\ \nu}\,F^\nu_{\ \lambda}\,F^\lambda_{\ \sigma}\,F^\sigma_{\ \mu}
		+16\, F_{\mu\nu}\,F^{\mu\lambda}\,F^{\nu\sigma}\,F_{\lambda\sigma} \right. \nonumber \\
		&-& \left. 4\, F_{\mu\nu}\,F^{\mu\nu}\,F_{\lambda\sigma}\,F^{\lambda\sigma}
		-2\, F_{\mu\nu}\,F_{\lambda\sigma}\,F^{\mu\nu}\,F^{\lambda\sigma}
	\right]\label{eq:SV2}
\end{eqnarray}
where we define
\begin{equation}
	F_{0 i} = \partial_t K^i\ \ \ ,\ \ \ F_{ij} = i\, [K^i,K^j]
\end{equation}
Notice that, given that the center of mass perturbations commute with all matrices, we have
\begin{equation}
	F_{ij} = i\, [\overline{K}^i,\overline{K}^j]\ .
\end{equation} 
And we also have 
\begin{equation}
	F_{0 i} = \partial_t \overline{K}^i + \partial_t (\varepsilon^i_1 - \varepsilon^i_2)\, \bm{1}_{N_1\times N_1} \otimes \bm{1}_{N_2\times N_2}\ .
\end{equation}
These time derivatives of $\varepsilon_{1,2}$ are sub-leading to the kinetic terms of the perturbations arising in~(\ref{eq:S02}) as they will be multiplied by $\sim r^{-7}$.
The terms involving $\partial_t{\varepsilon}_{1,2}$ can then be dropped as long as the distance between the two objects is large. We then get
\begin{equation}
	F_{ij}= i\, [\overline{K}^i,\overline{K}^j] = \overline{F}_{ij}\ \ \ ,\ \ \ F_{0 i} \simeq \partial_t \overline{K}^i = \overline{F}_{0 i}\ .
\end{equation}
Note next that the $\overline{F}_{ij}$ and $\overline{F}_{0 i}$ are independent of $r^i$, the separation vector between the two objects. To see this, we have from~(\ref{eq:kappa}) 
\begin{equation}
	\overline{F}_{ij} = i\, [\kappa^i,\kappa^j]
\end{equation}
where the matrix entries of $\kappa^i$ scale as the size of each object, independent of $r^i$. As for $\overline{F}_{0 i}$, we have from~(\ref{eq:kappa})
\begin{equation}
	\overline{F}_{0 i} = \partial_t r^i \bm{1}_{N_1\times N_1} \otimes \bm{1}_{N_2\times N_2} + \partial_t \kappa^i
\end{equation}
demonstrating that $\overline{F}_{0 i}$ is also $r^i$ independent -- but of course it depends on $\partial_t {r}^i$.
Putting things together, we can then write
\begin{equation}
	S_V \simeq  -\int dt\, \frac{1}{2} \varepsilon_a^i \varepsilon_b^j  \frac{\partial^2 V}{\partial \overline{x}_a^i \partial \overline{x}_b^j}
\end{equation}
where $a$ and $b$ sum over $1$ and $2$, and 
where $V$ is the potential from~\cite{Kabat:1997im}
\begin{equation}
	V = -\frac{5}{128\, r^7} W
\end{equation}
with
\begin{equation}
	W = \mbox{Tr} \left[
		8\, \overline{F}^\mu_{\ \nu}\,\overline{F}^\nu_{\ \lambda}\,\overline{F}^\lambda_{\ \sigma}\,\overline{F}^\sigma_{\ \mu}
		+16\, \overline{F}_{\mu\nu}\,\overline{F}^{\mu\lambda}\,\overline{F}^{\nu\sigma}\,\overline{F}_{\lambda\sigma} - 4\, \overline{F}_{\mu\nu}\,\overline{F}^{\mu\nu}\,\overline{F}_{\lambda\sigma}\,\overline{F}^{\lambda\sigma}
		-2\, \overline{F}_{\mu\nu}\,\overline{F}_{\lambda\sigma}\,\overline{F}^{\mu\nu}\,\overline{F}^{\lambda\sigma}
	\right]\ .
\end{equation}

Note that, as promised, we dropped terms linear in $\varepsilon$.
In~\cite{Kabat:1997im}, it was shown that $V$ matches precisely (including numerical coefficient) with the expected Newtonian gravitational potential averaged over the light-cone direction between the two objects as long as $N_{1,2}$ are large
\begin{equation}
	V = -\frac{15}{4} \frac{R^4}{N_1N_2\,r^7}\left((p_1\cdot p_2)^2-\frac{1}{9}p_1^2 p_2^2\right)
\end{equation}
where $p_1$ and $p_2$ are the eleven dimensional momenta of the two objects.

Combining this result with the rest of the action from~(\ref{eq:S02}), we then have the effective action for  $\varepsilon_1$ and $\varepsilon_2$ -- that represent diagonal perturbations of the two objects 
\begin{equation}\label{eq:effectiveS}
	\int dt\, \left(N_1 (\partial_t \varepsilon^i_1)^2 + N_2 (\partial_t \varepsilon^i_2)^2
	- \left(\frac{1}{2} \varepsilon_1^i \varepsilon_1^j
	+\frac{1}{2} \varepsilon_2^i \varepsilon_2^j
	-\varepsilon_1^i \varepsilon_2^j\right) \frac{\partial^2 V}{\partial r^i \partial r^j} \right)
\end{equation}
where we used the fact that, in the regime of large distance $r$ between the object, the potential $V$ depends only on $r^i=x_1^i-x_2^i$.

To proceed further, we need to setup a particular scenario where one of the two objects is treated as the heavy source and the other is a light probe. This sets the stage to interpreting the soon to be computed quantum entanglement as a measure of the local curved geometry experienced by the probe due to the source. Let us take object 1 to be the massive `star' whose geometry object 2 is probing; for example, we might write
\begin{equation}\label{eq:sphere}
	\overline{X}_1^i = \overline{x}_1^i\,\bm{1}_{N_1\times N_1} + \frac{2\,r_1}{N_1} J_i
\end{equation}
where the $J_i$ are the angular momentum matrices, satisfying the $SU(2)$ algebra and the Casimir relation
\begin{equation}
	[J_i,J_j] = i\,\epsilon_{ijk} J_k\ \ \ ,\ \ \ \mbox{Tr}\, J_i^2 \simeq \frac{N_1^3}{4}\ ,
 \end{equation}
where we assumed that $N_1\gg 1$. Similarly, we can take object 2 to be a spherical `planet' with $N_2$ units of light-cone momentum that is much lighter and smaller. Each object 
has a non-zero size $R_{1,2}$ which is, at the least, the radius of the corresponding black hole. However, spatially localized configurations like the one given by~(\ref{eq:sphere}) do not solve the equations of motion without an additional infrared cutoff -- {\em i.e.} we may not assume that the background is on shell as we have done so. If object 1 were to be a black hole, we expect that the chaotic nature of Matrix theory admits a metastable spherical configuration that is long-lived as it evaporates away slowly via Hawking radiation~\cite{Du:2018dmi}. It has been shown that this stochastic short timescale dynamics can be effectively modeled by adding by hand a quadratic mass term to the action. Alternatively, one can imagine a background flux that stabilizes the configuration like in the case of the giant gravitons of the Berenstein-Maldacena-Nastase (BMN) Matrix model~\cite{Berenstein:2002jq}. In either scenario, object 1 maintains a finite size due to some additional terms in the action, either due to effective stochastic physics or due to a non-flat background that essentially puts the system in a box. Here, we account for this by adding by hand a generic stabilizing term, the simplest of which would be
\begin{equation}
	S\rightarrow S - \int dt\ \alpha_1\, \left(\mbox{Tr} {X}_1^i{X}_1^i - \frac{\mbox{Tr}\,X_1^i\ \mbox{Tr}\,X_1^i}{N_1}\right)
	- \int dt\ \alpha_2\, \left(\mbox{Tr} {X}_2^i{X}_2^i - \frac{\mbox{Tr}\,X_2^i\ \mbox{Tr}\,X_2^i}{N_2}\right)
\end{equation} 
where $\alpha_{1,2}$ are positive constants that are tuned to assure a given stable sizes $R_{1,2}$ for objects 1 and 2\ \footnote{For example, it is easy to check that, for a spherical configuration of radius $R_1$ given by~(\ref{eq:sphere}), one needs $\alpha_1 = 8\,R_1^2/N_1^2$.}. The important general observation is that $\alpha_{1}$ and $\alpha_2$ must be positive to assure stability, and they are larger for larger objects.
To see this, for the configuration given by~(\ref{eq:sphere}), we can check that the size of object 1 is $R_1=r_1$, and its mass scales as $M\sim r_1^2 \sim \alpha_1\,N_1^2$ (the area of the spherical membrane). For fixed light-cone momentum $N_1$, large $\alpha_1$ corresponds to larger energy. 
Treating object 2 as the light probe, we henceforth assume that $\alpha_2\ll \alpha_1$. In fact, as we shall see, it does not matter which one of the two objects is the lighter probe -- the entanglement entropy of either one is the same as the other's, as expected from the fact that the combined system of diagonal perturbations is in a pure state. 

The result of this is that one ends up adding an additional terms to the effective action~(\ref{eq:effectiveS}) of the form $-N_1\, \alpha_1\, (\varepsilon_1^i)^2$ and $-N_2\, \alpha_2\, (\varepsilon_2^i)^2$ which dominate the corresponding $(\varepsilon_1^i)^2$ and $(\varepsilon_2^i)^2$ terms in~(\ref{eq:effectiveS}). We then have the modified effective action
\begin{equation}\label{eq:effectiveS2}
	\int dt\, \left(N_1 (\partial_t \varepsilon^i_1)^2 + N_2 (\partial_t \varepsilon^i_2)^2 - N_1\, \alpha_1\, (\varepsilon_1^i)^2- N_2\, \alpha_2\, (\varepsilon_2^i)^2
	+\varepsilon_1^i \varepsilon_2^j \frac{\partial^2 V}{\partial r^i \partial r^j}\right)\ .
\end{equation}
We rescale the perturbations so as to canonically normalize the kinetic terms 
\begin{equation}
	z_{1,2}\equiv \sqrt{N_{1,2}}\,r\,\varepsilon_{1,2}\ .
\end{equation}
We end up with the final effective action for the perturbations\footnote{One can also consider the probe to be a graviton. As a result,  $\alpha_2\rightarrow 0$ and we must keep the $\varepsilon_2^2$ term from~(\ref{eq:effectiveS}). The subsequent computation is then slightly modified and the general pattern persists as long as the probe is much lighter than the source.}
\begin{equation}\label{eq:effectiveS3}
	S_{eff} = \int dt\, \left(
	(\partial_t z_1)^2 + (\partial_t z_2)^2 - \alpha_1\, z_1^2 - \alpha_2\, z_2^2 
	+\frac{z_1^i z_2^j}{\sqrt{N_1N_2}}\, \partial_i \partial_j V
	\right)\ .
\end{equation}
We write $\partial_i = \partial / \partial z_2^i$, derivatives with respect to the probe's location.
This is the effective action that describes the diagonal perturbations, to leading order $R_1/r$ and $R_2/r$, between blocks 1 and 2 of the matrices -- in a regime where object 2 is a light probe under the influence of a massive object 1 that curves the spacetime around it. We next compute the quantum entanglement in the vacuum of the $z_1$-$z_2$ system arising from the $z_1 z_2$ coupling term in this effective action. 
 
We have a system with two degrees of freedom with a Hamiltonian
\begin{equation}
	H = (\partial_t z_1)^2 + (\partial_t z_2)^2 + \alpha_1\, z_1^2+ \alpha_2\, z_2^2
	- \frac{1}{\sqrt{N_1N_2}} z_1^i z_2^j\, \partial_i \partial_j V
	 \equiv (\partial_t z_{a})^2 + z^i_a W_{(ai)(bj)} z^j_b \ ,
\end{equation}
where $a,b$ sum over $1,2$.
Following~\cite{Casini:2009sr}, we define the matrix $\omega$ as
 \begin{equation}
 	\omega = W^{1/2}\ \ \ \mbox{where}\ \ \ W = \left(\begin{array}{cc}\alpha_1\,\delta_{kl} & -\frac{1}{2\sqrt{N_1N_2}} \partial_i\partial_k V \\ -\frac{1}{2\sqrt{N_1N_2}} \partial_l\partial_j V &  \alpha_2\,\delta_{ij}\end{array}\right)\ ,
 \end{equation}
 in $2\times 2$ block diagonal form.
The density matrix for the vacuum state takes the form
 \begin{equation}
 	\rho(z,z') = \sqrt{\frac{\mbox{det}\,\omega}{\pi}} e^{-\frac{1}{2}z^T\omega\, z}e^{-\frac{1}{2}{z'}^T\omega\, z'}\ .
 \end{equation}
In our case, we get
\begin{equation}
	\omega = \left(\begin{array}{cc}
	\sqrt{\alpha_1} & -\frac{1}{2\sqrt{\alpha_1}\sqrt{N_1N_2}} \partial_i\partial_k V \\
	-\frac{1}{2\sqrt{\alpha_1}\sqrt{N_1N_2}} \partial_l\partial_j V & \sqrt{\alpha_2}
	\end{array}\right)\ ,
\end{equation}
where we have evaluated the square root of the matrix in the regime where (a) the off-diagonal entries of $W$ are much smaller than the diagonal
ones; and where (b) we have $\alpha_2 \ll \alpha_1$ since object 2 is the probe.  

We are interested in computing the entanglement entropy of object 2 with object 1 by tracing over the Hilbert space of object 1 and computing the Von Neumann entropy of the resulting reduced density matrix. Following~\cite{Casini:2009sr}, we then define
 \begin{equation}\label{eq:lambda}
 	\Lambda \equiv \omega_{22}^{-1/2}\cdot\omega_{21}\cdot\omega_{11}^{-1}\cdot\omega_{12}\cdot\omega_{22}^{-1/2}
 \end{equation}
where $\omega_{11}$ is the sub-block of the matrix on the diagonal referring to object 1, $\omega_{12}$ is the sub-block between objects 1 and 2, etc... $\Lambda$ is then a $9\times 9$ matrix in the tangent space of the probe's location -- parameterized by $z_2^i$. For our case, we have
\begin{equation}\label{eq:Lambda}
	\Lambda_{ij} = \frac{1}{4\,\alpha_1\,N_1 N_2} \frac{(\partial_i \partial_k V)(\partial_k \partial_j V)}{\sqrt{\alpha_1\alpha_2}}\equiv \gamma^2\, (\partial_i \partial_k V)(\partial_k \partial_j V)\ .
\end{equation}
We have defined $\gamma$ to absorb all constants that refer to information about the individual objects, such as their sizes, masses, and equations of state. Note also that the eigenvalues of $\Lambda$ are much smaller than one in the regime we have been working in.

The Von Neumann entropy of interest is then given by~\cite{Casini:2009sr}
\begin{eqnarray}
	S_{ent}(\Lambda) &=& \mbox{Tr} \left(
	\ln \frac{1-\Lambda/2+\sqrt{1-\Lambda}}{1-\Lambda+\sqrt{1-\Lambda}}
	- \frac{\Lambda}{2} \frac{\ln \frac{\Lambda}{2-\Lambda+2\sqrt{1-\Lambda}}}{1-\Lambda+\sqrt{1-\Lambda}}
	\right) \nonumber \\
	&\simeq & - \mbox{Tr} \left( \frac{\Lambda}{4} \ln \frac{\Lambda}{4}\right)+ \mbox{Tr}\, \frac{\Lambda}{4}\ ,\label{eq:sent}
\end{eqnarray}
where the simpler form on the second line is valid when the eigenvalues of $\Lambda$ are much smaller than one, as is the case for us. Note than all $U(N)$ matrix structure has disappeared and the relevant object lives in the tangent space of the probe's position -- the vector space over which the expression traces. We have hence computed the entanglement entropy of the two objects in the quantum vacuum of perturbations of their centers of mass, and have shown how this entropy is a function of the gravitational potential that the probe experiences due to the presence of the source. 

\section{A new entropy-geometry relation}\label{sec:area}

The gravitational potential $V$ encodes information about the curvature of the spacetime at the probe's location. This means that we must be able to relate the entanglement entropy of the probe to local spacetime geometry. We start on the general relativity side with the light-cone gauge M-theory probe evolving along a timelike geodesic with tangent denoted by $u^\mu$, where $\mu=0,1,\cdots, 10$. Indices $1,\cdots, 9$ are the transverse directions to the light-cone, mapping onto the Matrix theory target space indices; while the theory is boosted in light-cone direction $x^{10}$. Let $z_{i}^\mu$ be nine spacelike vectors tangent to $u^\mu$, so we have $i,j=1,\cdots, 9$. One can project on this sub-space using
\begin{equation}
	h^{\ \mu}_{\nu} = \delta^\mu_\nu+u^\mu u_\nu\ .
\end{equation}
We can then relate the Newtonian gravitational potential $V$ of the probe to the local Riemann tensor that it samples by~\cite{Hawking:1973uf}
\begin{equation}\label{eq:VR}
	z_i^\mu z_j^\nu V_{;\mu\nu} \simeq z_i^\mu z_j^\nu R_{\mu\rho\nu\sigma} u^\rho u^\sigma\ .
\end{equation}
Looking back at~(\ref{eq:Lambda}), we see that the entropy is expressed as a function of the double derivatives of the potential, instead of covariant derivatives. This is natural in the context of Matrix theory as the Matrix theory formulation is background dependent, built up on top of a flat Minkowski background. This suggests that the probe coordinates on the Matrix theory side of the correspondence cannot map onto general coordinates that the dual M-theory geometry might be written in. We then conjecture that one is required to interpret the Matrix theory coordinates as {\em locally flat} coordinates at the location of the probe on the M-theory side\footnote{Note that~(\ref{eq:VR}) does {\em not} map onto the desired form involving simple derivatives at  asymptotic infinity where curvatures are weak and where our computation is designed to hold. Hence, there is no alternative to locally flat coordinates, where the Christoffel symbols vanish. Note also that this is a more general coordinate system than Riemann normal or Fermi normal coordinates, and there is still infinite freedom {\em globally} in fixing locally flat coordinates. At the location of the probe, the freedom consists of local rotations $SO(9)$, a subgroup of the gauge group of eleven dimensioanl gravity given the resriction to light-cone gauge. As required, this is also the symmetry group on the Matrix theory side.}. Matrix theory would then build up geometry locally through probe tidal acceleration that the Matrix effective potential can naturally determine. In locally flat coordinates at the location of the probe, the Christoffel symbols vanish and we have
\begin{equation}\label{eq:zz}
	z_i^\mu z_j^\nu  V_{,\mu\nu} = z_i^\mu z_j^\nu \partial_\mu \partial_\nu V = z_i^\mu z_j^\nu R_{\mu\rho\nu\sigma} u^\rho u^\sigma\ .
\end{equation}
We also have at the location of the probe $\eta_{\mu\nu}z_{i}^\mu u^\nu=0$ for $i=1,\cdots,9$. It is easy to check that we can write
\begin{equation}
	z_i^\mu = \delta^\mu_i-\frac{u_i}{u_-} \delta^\mu_-\ ,
\end{equation}
where we use the light-cone metric such that $-2\,u^+u^-+(u^i)^2=-1$. Note that the light-cone momentum $p^+ = N_2/R$, and the light-cone energy is $p^- = (m^2+(p^i)^2)/(2\,p^+)$. We also have
$\partial_- V =0$ from the fact that $V$ is averaged over $x^-$ since no longitudinal momentum is exchanged between source and probe. We then can write
\begin{equation}
	\partial_i \partial_j V = \left(\delta^\mu_i-\frac{u_i}{u_-} \delta^\mu_-\right)\left(\delta^\nu_j-\frac{u_j}{u_-} \delta^\nu_-\right)R_{\mu\rho\nu\sigma} u^\rho u^\sigma \equiv \mathcal{R}_{ij}\ ,
\end{equation}
defining the new quantity $\mathcal{R}_{ij}$ built out of the local Riemann tensor, or equivalently tidal forces.

Putting things together, we write
\begin{equation}\label{eq:main}
	S_{ent} \simeq - \gamma^2\,\mbox{Tr} \left( \frac{\mathcal{R}^2}{4} \ln \frac{\mathcal{R}^2}{4}\right)\ .
\end{equation}
This is a local relation between the curvature sampled by the probe and the quantum entanglement between the center of mass degrees of freedom of source and probe. Note that this entropy is finite, not surprisingly given that we are working in a UV complete theory of quantum gravity. In the limit where the curvature vanishes, so that this expression for entropy.
Next, we consider the expression
\begin{equation}
	\theta_{\mu\nu} = h^{\ \alpha}_{\mu} h^{\ \beta}_{\nu} u_{(\alpha;\beta)}
\end{equation}
which is a measure of deformations of the shape and orientation of a small sphere at the probe along its trajectory. We then have a version of Raychaudhuri's equation
\begin{equation}
	\frac{d}{d\tau} \theta_{\mu\nu} \simeq - R_{\mu\rho\nu\sigma} u^\rho u^\sigma
\end{equation}
where we have dropped higher order terms that are smaller than the leading contribution at weak curvatures. Using locally flat coordinates, and projecting onto the nine dimension subspace using the $z_i^\mu$'s, we have
\begin{equation}
	\frac{d}{d\tau} \theta_{ij} \simeq - z^\mu_i z^\nu_j R_{\mu\rho\nu\sigma} u^\rho u^\sigma = -\mathcal{R}_{ij}\ .
\end{equation}
In particular $d\theta/d\tau$, where $\theta$ is the trace of $\theta_{ij}$ using the metric $h_{ij}$, is the rate of change of a volume element along the probe's geodesic. Hence, equation~(\ref{eq:main}) establishes a relation between the source-probe entanglement entropy and the rate at which a small volume of space shrinks, rotates, and twists along the geodesic of the probe. If we were to choose the probe to be massless, one can easily show that one obtains a similar relation but now involving an {\em area} transverse to a congruence of {\em null} geodesics associated with the probe. All this is somewhat reminiscent of the entropy-area relations we encounter in other settings~\cite{Ryu:2006bv,Nishioka:2009un} with one significant difference being that our relation is a {\em local} statement.

\section{Conclusion}\label{sec:conclusion}

We have demonstrated that Von Neumann entanglement entropy between two blocks of matrices in Matrix theory -- that represent a probe gravitating near a source -- can quite generically be written as a function of derivatives of their mutual gravitational potential. We also presented  arguments and a conjecture for expressing this relation as a map between entanglement entropy and local spacetime geometry as sampled by the probe in the background of the source. 

We considered a particular scenario and worked consistently only to leading order in weak gravitational potential energy. Yet, the analysis introduces a new general way to develop maps between quantum information and spacetime geometry in Matrix theory. This involves looking at diagonal matrix fluctuations and focusing on the ground state density matrix of these degrees of freedom. As a result, when one focuses on a sub-block of a matrix, the resulting reduced density matrix and quantum entanglement will be related to the effective Matrix potential between the two matrix sub-blocks arising from integrating out fast off-diagonal modes. This mechanism appears general and might hint to why, at least to leading order in weak gravity, one expects a relation between entanglement entropy and spacetime geometry. 

Entanglement entropy by nature is multi-faceted. It depends on how one slices parts of a larger system, and on what quantum state the entire system lives in. These freedoms are very much reflected in the analysis, where we made a series of choices to set a computationally accessible setup. There are many more settings to explore, and a catalogue of case studies can help develop intuition on the general pattern of expected relations between entropy and local geometry. We end by pointing out a couple of particularly interesting cases: the case involving massless probes, where one has the promise to connect with ideas from holography and entropy of light-sheets developed from different perspectives~\cite{Bousso:1999xy,Sahakian:1999bd,vvs}; and the case where the approach is used in BMN theory that admits stable giant gravitons and hence the need to add stablizing terms to the action is avoided~\cite{vvs2}.

\newpage
\section{Acknowledgments}

This work was supported by NSF grant number PHY-0968726.

\providecommand{\href}[2]{#2}\begingroup\raggedright\endgroup


\end{document}